\documentclass[aps,pra,floatfix,twocolumn,superscriptaddress]{revtex4-1}

\usepackage{graphicx}
\usepackage{dcolumn}
\usepackage{bm}

\usepackage{color}

\usepackage{xcolor}
\usepackage{soul}
\usepackage{graphicx}
\usepackage{bm}
\usepackage{mathrsfs}
\usepackage{amsmath}
\def\lesssim{\ \raise.3ex\hbox{$<$}\kern-0.8em\lower.7ex\hbox{$\sim$}\ }
\def\gesim{\ \raise.3ex\hbox{$>$}\kern-0.8em\lower.7ex\hbox{$\sim$}\ }

\begin{document}
\title{{Thermal liquid-gas phase transition in a quasi-one-dimensional dipolar Fermi gas}}

\author{Lanxuan Gao}
\thanks{These authors contributed equally.}
\email{gao-lanxuan102@g.ecc.u-tokyo.ac.jp}
\affiliation{Department of Physics, School of Science, The University of Tokyo, Tokyo 113-0033, Japan}

\author{Koki Takayama}
\thanks{These authors contributed equally.}
\affiliation{Department of Physics, School of Science, The University of Tokyo, Tokyo 113-0033, Japan}

\author{Hiroyuki Tajima}
\affiliation{Department of Physics, Graduate School of Science, The University of Tokyo, Tokyo 113-0033, Japan}
\affiliation{RIKEN Nishina Center, Wako 351-0198, Japan}
\affiliation{Quark Nuclear Science Institute, The University of Tokyo, Tokyo 113-0033, Japan}

\author{Takahiro M.\ Doi}
\affiliation{Department of Physics, Kyoto University, Kitashirakawa Oiwakecho, Sakyo-ku, Kyoto, 606-8502, Japan}
\affiliation{
    RIKEN Center for Interdisciplinary Theoretical and Mathematical Sciences (iTHEMS),
    Wako 351-0198, Japan}

\author{Haozhao Liang}
\affiliation{Department of Physics, Graduate School of Science, The University of Tokyo, Tokyo 113-0033, Japan}
\affiliation{Quark Nuclear Science Institute, The University of Tokyo, Tokyo 113-0033, Japan}
\affiliation{
    RIKEN Center for Interdisciplinary Theoretical and Mathematical Sciences (iTHEMS),
    Wako 351-0198, Japan}

\date{\today}
\begin{abstract}
We theoretically investigate the thermodynamic properties of a quasi-one-dimensional single-component dipolar Fermi gas at finite temperatures.
A self-bound fermionic droplet can be achieved by exchange correlations with long-range dipole-dipole interactions under quasi-one-dimensional confinement, where the interaction can be tuned by tilting the dipoles along the system coordinate.
Using the Hartree-Fock approximation, we show how the liquid-gas phase transition occurs in this system and elucidate the finite-temperature phase structure consisting of a gas phase, liquid phase, gas-liquid coexistence phase, and spinodal phase.
We also discuss its similarity to the liquid-gas phase transition in nuclear matter through a comparison with phenomenological models.
By examining the experimental conditions for realizing self-bound fermionic droplets, we find that microwave-shielded fermionic polar molecules are promising candidates.
Our results will be useful for an interdisciplinary understanding of self-bound fermionic matter as well as an analog quantum simulation of nuclear systems.
\end{abstract}

\maketitle

\section{Introduction}

A liquid-gas phase transition is a ubiquitous phenomenon in many-body systems.
The transition of classical liquids such as water and oil has been studied for a long time.
Typically, it can be described by the van der Waals equation of state with a short-range repulsive interaction and a long-range attractive interaction.

In modern physics, the liquid-gas phase transition in quantum many-body systems has attracted much attention.
In condensed-matter physics, it is known that excitons with a strong dipole-dipole interaction exhibit
a first-order phase transition from the gas phase to the liquid phase with increasing the carrier density~\cite{PhysRevB.7.1508, keldysh1986electron, PhysRevLett.120.047402}.
Another illuminating example in quantum many-body systems is atomic nuclei consisting of nucleons with strong nuclear force, where spin and isospin degrees of freedom and non-local nucleon-nucleon interactions involving the so-called tensor force play a crucial role~\cite{BORDERIE201982}. In particular, the nuclear matter equation of state, responsible for the formation of self-bound nuclei, is a central issue in nuclear physics, and moreover, it is deeply related to neutron star physics~\cite{lattimer2021neutron}.

Recently, the self-bound quantum liquid droplet has also been discussed in ultracold Bose gases~\cite{PhysRevLett.115.155302},
where quantum fluctuations suppress the mean-field collapse.
This state has been experimentally realized
in a dipolar Bose atomic gas~\cite{PhysRevLett.116.215301} and molecules~\cite{zhang2025observation}
as well as in
a binary Bose-Bose mixture~\cite{cabrera2018quantum,PhysRevLett.120.235301}.
In this regard, one can study quantum droplets and liquids with tunable ultracold gases, which can be regarded as an analog quantum simulator of quantum many-body systems.
Accordingly, thermal destabilization and evaporation of quantum droplets have been discussed theoretically~\cite{wang2020thermal,PhysRevA.103.043316,PhysRevLett.131.173404}.

However, its fermionic counterpart has not yet been observed in cold atomic physics. The realization of self-bound droplets in fermionic systems is a promising candidate for simulating nuclei and nuclear matter, and would be a useful testing ground for nuclear density functional theories~\cite{kemler2016formation,liang2018functional,PhysRevC.99.024302}.
Theoretically, it is reported that single-component dipolar fermions in a quasi-one-dimensional geometry
can form the self-bound state~\cite{PhysRevA.88.033611}, where the dipole-dipole interaction can be tuned to be attractive by tilting the dipoles along the system geometry~\cite{PhysRevLett.99.140406}.
Moreover, the quasi-one-dimensional geometry enables the system to avoid the mechanical collapse due to the long-range attractive interaction.
Such a system can be realized in cold atom experiments with dipolar Fermi atoms~\cite{PhysRevLett.108.215301} and polar molecules~\cite{de2019degenerate}.

\begin{figure}[t]
    \centering
    \includegraphics[width=0.9\linewidth]{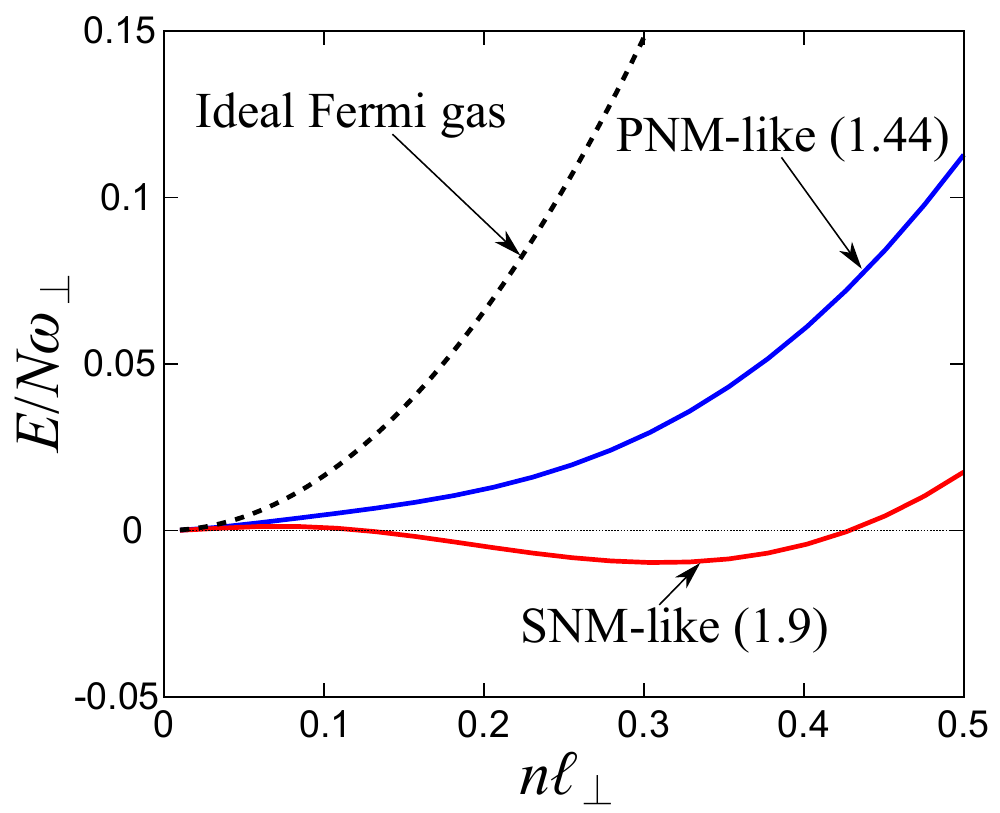}
    \caption{
    {Zero-temperature energy per particle $E/N\omega_\perp$ in a quasi-one-dimensional dipolar Fermi gas at the gas parameter $\alpha \ell_{dd}/\ell_\perp=1.44$ and $\alpha\ell_{dd}/\ell_\perp=1.90$, obtained by the Hartree-Fock approximation~\cite{PhysRevA.88.033611}. The horizontal axis is the number density $n=N/L$ where $N$ and $L$ are the particle number and the system length. 
    $\omega_\perp$ and $\ell_\perp$ are the transverse harmonic trap frequency and associated length scale, respectively. 
    $\alpha$ is a dimensionless quantity for the dipolar angle.
    The dotted line shows the result of an ideal Fermi gas.
    While the case of $\alpha \ell_{dd}/\ell_\perp=1.44$ corresponds to the threshold for a two-body bound state and hence can be regarded as an analog of pure neutron matter (PNM), the case of $\alpha\ell_{dd}/\ell_\perp=1.90$ involving the two-body bound state may be analogous to symmetric nuclear matter (SNM) with a local minimum of $E/N$.}}
    \label{fig:1}
\end{figure}

It is worth mentioning that there is a critical interaction strength (dipole angle) for the two-body bound state in a quasi-one-dimensional dipolar Fermi gas~\cite{PhysRevA.88.033611}.
It is similar to the $p$-wave unitarity in one dimension~\cite{PhysRevA.94.043636,PhysRevA.104.023319}, and reminiscent of the similarity between a unitary Fermi gas and dilute pure neutron matter~\cite{PhysRevA.63.043606,PhysRevC.77.032801,PhysRevA.97.013601}.
In addition, one can prepare the self-bound fermionic droplet by increasing the attractive interaction whose equation of state (EOS) shares similarities with that in symmetric nuclear matter.
{In nuclear physics, tremendous efforts have been made to extract the thermodynamic quantities of symmetric nuclear matter, such as incompressibility and symmetry energy, from the experimental observables of heavy ion collisions as well as astrophysical data~\cite{hebeler2013equation}.
However, these extractions involve theoretical assumptions to some extent.}
In this regard, one can investigate a cold-atomic system exhibiting EOS being similar to pure neutron matter and symmetric nuclear matter within a common model, as shown in Fig.~\ref{fig:1},
{which enables us to microscopically quantify the theoretical uncertainties associated with the connection between experimental observables and EOS in self-bound fermionic systems.}

For future experimental realizations, it is important to investigate the finite-temperature effect as the experiments are always conducted at finite temperatures. 
The thermal EOS is also important for understanding heavy-ion collisions~\cite{BORDERIE201982} and astrophysical phenomena such as core-collapsed supernova and neutron-star merger~\cite{RevModPhys.89.015007}.
{In particular, the finite temperature effect induces the thermal liquid-gas phase transition, where the self-bound liquid state turns into the gas phase under the competition between the kinetic and interaction energies. Such a phase transition has been discussed in nuclear matter within the Hartree-Fock approximation~\cite{Rios2010phase}.}

In this paper, we theoretically investigate the formation of the self-bound liquid state and its thermal phase transition in a quasi-one-dimensional dipolar Fermi gas. 
To this end, we employ the Hartree-Fock approximation, which can describe the formation of the self-bound state at zero temperature~\cite{PhysRevA.88.033611}. 
The Hartree-Fock approximation is also used for the description of three-dimensional dipolar Fermi gas at finite temperature~\cite{PhysRevA.81.033617}.
Extending the formulation for the quasi-one-dimensional self-bound state to the finite-temperature case, we show the existence of a liquid-gas phase transition and a spinodal phase in such a dipolar Fermi gas at finite temperature in the thermodynamic limit.

{We note that, in one dimension, low-energy properties of interacting fermions are generically described by Tomonaga-Luttinger liquid (TLL) theory rather than Fermi-liquid theory~\cite{RevModPhys.84.1253}.
In particular, pairing long-range order in one-dimensional spinless fermions with the weakly attractive dipole-dipole interaction has been discussed within the TLL framework at low temperature~\cite{Yan_2014}.
Meanwhile, both the TLL and Hartree-Fock frameworks can describe the spinodal instability (i.e., divergence of the static compressibility).
In the present work, the Hartree-Fock approximation is not used for the description of low-energy collective excitations, but for a qualitative understanding of the EOS in the self-bound fermionic system.}

This paper is organized as follows. In Sec.~\ref{sec:2}, we show the theoretical model and the formulation of the Hartree-Fock theory in a quasi-one-dimensional dipolar Fermi gas. In Sec.~\ref{sec:3}, we present our numerical results of the liquid-gas phase transition and the finite-temperature phase diagram.
Moreover, we discuss the comparison with phenomenological models.
Finally, we summarize this paper in Sec.~\ref{sec:4}. For simplicity, we use a unit of $\hbar=k_{\rm B}=1$ and the system length $L$ {can be} taken to be unity in the thermodynamic limit {due to its homogeneity}.

\section{Formalism}\label{sec:2}

\subsection{Model}

We consider quasi-one-dimensional dipolar fermions with the dipole moment $d$  described by the Hamiltonian 
\begin{align}
    H=\sum_{k}\xi_kc_k^\dag c_k
    +\frac{1}{2{L}}\sum_{k,k',q}V_{dd}(q)c_{k}^\dag c_{k'+q}^\dag c_{k'}c_{k+q},
\end{align}
where $\xi_k=k^2/2m-\mu$ is the kinetic energy with a momentum $k$ and a mass $m$ measured from the chemical potential $\mu$.
$c_{k}$ and $c_{k}^{\dag}$ represent annihilation and creation operators of a single-component dipolar fermion.
The dipole-dipole interaction $V_{dd}(q)$ in the quasi-one-dimensional geometry reads~\cite{PhysRevLett.89.130401,PhysRevLett.99.140406}
\begin{align}
\label{eq:2}
    V_{dd}(q)=-\frac{2\alpha \ell_{dd}}{m\ell_\perp^2}[1-\sigma_\perp e^{\sigma_\perp}\Gamma(0,\sigma_\perp)],
\end{align}
where $\sigma_\perp=q^2\ell_\perp^2/2$ is a dimensionless squared momentum
with the transverse harmonic-oscillator length $\ell_\perp=1/\sqrt{m\omega_\perp}$ ($\omega_\perp$ is the transverse trap frequency).
$\ell_{dd}=md^2$ is a length scale associated with the dipole-dipole interaction.
Also, in Eq.~\eqref{eq:2}
\begin{align}
    \Gamma(a,x)=\int_x^{\infty}dt\, t^{a-1}{e^{-t}}
\end{align}
is an incomplete gamma function, and
\begin{align}
    \alpha=\frac{3\cos^2\varphi-1}{2},
\end{align}
represents the angle of dipoles $\varphi$ ($0\leq\varphi\leq\pi/2$) with respect to the longitudinal axis, satisfying
\begin{align}
    -\frac{1}{2}\leq\alpha\leq 1,
\end{align}
where $\alpha >0 \,(<0)$ is attractive (repulsive).
In particular, $\alpha=-1/2$ ($\varphi=\pi/2$) and $\alpha=1$ ($\varphi=0$).
Furthermore, 
$\alpha=0$ ($\varphi\approx 54.7^\circ$) is called a magic angle where the dipolar interaction disappears~\cite{PhysRevLett.99.140406}. {The function $[1-\sigma e^\sigma\gamma(0,\sigma)]$ in Eq.~\eqref{eq:2} is an even function with respect to $q\ell_\perp$ with the value of $1$ at $q\ell_\perp=0$. It asymptotically  behaves as $1-\sigma_\perp e^{\sigma_\perp}(-\gamma_E-\ln\sigma_\perp)$ at $q\ell_\perp \ll 0$ and $1/\sigma_\perp$ at $q\ell_\perp \gg 1$, where $\gamma_E$ is the Euler constant. The shape is plotted in Fig.~\ref{fig:1.5}.}

{
\begin{figure}[t]
    \centering
    \includegraphics[width=0.95\linewidth]{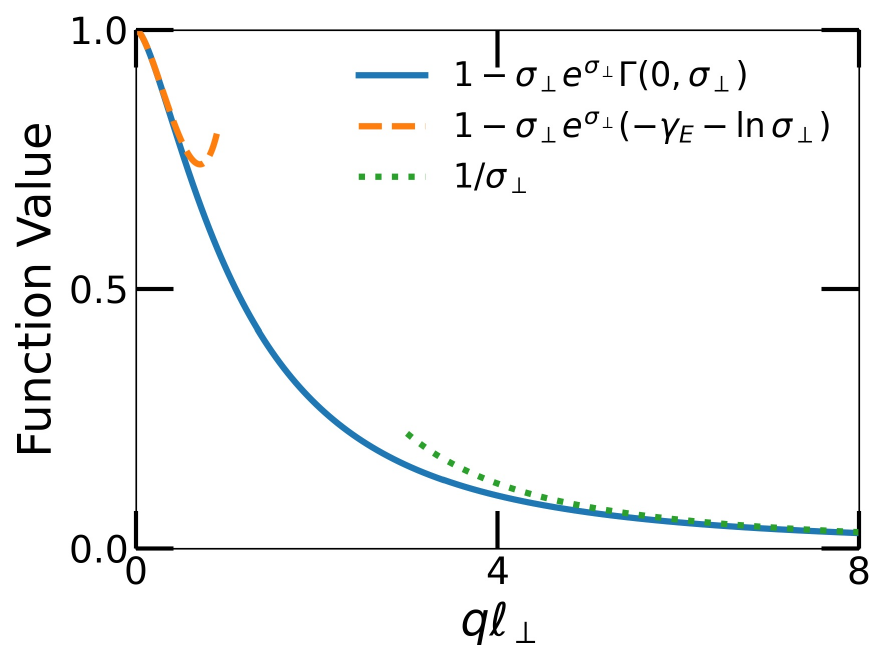}
    \caption{
    Plot of the function $[1-\sigma_\perp e^\sigma_\perp\gamma(0,\sigma_\perp)]$ in Eq.~\eqref{eq:2} as a function of $q\ell_\perp$. The function is even with respect to $q\ell_\perp$ with the value of $1$ at $q\ell_\perp=0$. It asymptotically  behaves as $1-\sigma_\perp e^{\sigma_\perp}(-\gamma_E-\ln\sigma_\perp)$ at $q\ell_\perp \ll 0$ and $1/\sigma_\perp$ at $q\ell_\perp \gg 1$.}
    \label{fig:1.5}
\end{figure}}

\subsection{Hartree-Fock approximation}
In this paper, we employ the Hartree-Fock approximation at finite temperature.
Although the Hartree-Fock approximation gives a larger ground-state energy density compared to the results based on the Bose-Fermi duality assuming that the quasi-one-dimensional dipole-dipole interaction can be described by the short-range interaction~\cite{PhysRevA.88.033611},
the Hartree-Fock result qualitatively agrees with this duality-based calculation in the entire density regime and, moreover, gives an accurate evaluation of the thermodynamic quantities at higher densities.  

Using the mean field $\langle c_{k}^\dag c_{k}\rangle$, one can reduce the interaction term into the one-body potential as
\begin{align}
    c_{k}^\dag c_{k}
    c_{p}^\dag c_{p}\approx
    \langle c_{k}^\dag c_{k}\rangle c_{p}^\dag c_{p}
    +
    \langle c_{p}^\dag c_{p}\rangle
    c_{k}^\dag c_{k}
    -
    \langle c_{k}^\dag c_{k}\rangle
    \langle c_{p}^\dag c_{p}\rangle.
\end{align}
In this regard,
we introduce the Hartree-Fock self-energy 
\begin{align}
    \Sigma_{k}^{\rm HF}
    &=
    {\frac{1}{L}}\sum_{{k}'}
    V_{dd}({0})
    \langle c_{k'}^\dag c_{k'}\rangle
    -{\frac{1}{L}}\sum_{{q}}
    V_{dd}({q})
    \langle c_{k+q}^\dag c_{k+q}\rangle,
\end{align}
where 
\begin{align}
    \langle c_{k}^\dag c_{k}\rangle = f(\xi_{k}+\Sigma_{k}^{\rm HF})\equiv \frac{1}{e^{\beta (\xi_{k}+\Sigma_k^{\rm HF})}+1}
\end{align}
is the momentum distribution with the Fermi-Dirac distribution function $f(x)=1/(e^{\beta x}+1)$ ($\beta=1/T$ is the inverse temperature).
Eventually, the mean-field Hamiltonian reads
\begin{align}
\label{eq:9}
    H_{\rm MF}=\sum_{{k}}
    \left(\xi_{{k}}+\Sigma_{{k}}^{\rm HF}\right)c_{{k}}^\dag c_{{k}}
    -\frac{1}{2}\sum_{{k}}
    \Sigma_{{k}}^{\rm HF}\langle c_{{k}}^\dag c_{{k}}\rangle
    .
\end{align}
Accordingly, the internal energy is given by
\begin{align}
E&=\langle H_{\rm MF}+\mu N\rangle\cr
&=\sum_{k}\left(\frac{k^2}{2m}+\frac{1}{2}\Sigma_{k}^{\rm HF}\right)f(\xi_{k}+\Sigma_{k}^{\rm HF}). \label{eq:internal_energy}
\end{align}

To obtain thermodynamic quantities at given $N$, we need to solve the number-density equation
\begin{align}
\label{eq:10}
    N=\sum_{k}f(\xi_{k}+\Sigma_{k}^{\rm HF}), 
\end{align}
noting that the particle number density $n$ is equal to $N$ in units of $L=1$.
From Eq.~\eqref{eq:10}, one can numerically obtain $\mu$ at a fixed $N$.
$\Sigma_{k}^{\rm HF}$ is also obtained by numerically solving the self-consistent equation
\begin{align}
\label{eq:12}
    \Sigma_{k}^{\rm HF}
    &=
    {\frac{1}{L}}\sum_{{k}'}
    V_{dd}({0})
    f(\xi_{{k}'}+\Sigma_{{k}'}^{\rm HF})\cr
    &\quad -{\frac{1}{L}}\sum_{{q}}
    V_{dd}({q})f(\xi_{{k}+{q}}+\Sigma_{{k}+{q}}^{\rm HF}).
\end{align}

{In the thermodynamic limit,} the pressure $P$ can be obtained from the thermodynamic potential $\Omega$ as
\begin{align}
    P=-{\frac{1}{L}}\Omega
    =\frac{1}{\beta{L}}\ln\Xi,
\end{align}
where
$\Xi\equiv \mathrm{tr}[e^{-\beta H_{\mathrm{MF}}}]$
is the grand-canonical partition function within the Hartree-Fock approximation.
Using Eq.~\eqref{eq:9}, one can obtain
\begin{align}
\label{eq:14}
\begin{split}
P &=-{\frac{1}{L}}
\sum_{k}
\,\Bigg[
 \tfrac12\,\Sigma_k^{\rm HF} f(\xi_k+\Sigma_k^{\rm HF})  \\
&\qquad\qquad +\, \frac{1}{\beta}\ln\!\Bigl(1+e^{-\beta(\xi_k+\Sigma_k^{\rm HF})}\Bigr)
\Bigg].
\end{split}
\end{align}
In this way, one can investigate $P$ as a function of {the particle density} $n\,{\equiv N/L}$ and $T$.

At finite temperature, we discuss the liquid-gas phase transition of the present system.
For this purpose, it is important to find the so-called flash point and the critical point~\cite{Rios2010phase} defined by
\begin{align}
\label{eq:15}
    P=\left(\frac{\partial P}{\partial n}\right)_T=0 \quad{\rm (flash \ point)},
\end{align}
and 
\begin{align}
\label{eq:16}
    \left(\frac{\partial P}{\partial n}\right)_T=\left(\frac{\partial^2 P}{\partial n^2}\right)_T=0 \quad{\rm (critical \ point)},
\end{align}
respectively.
In particular, we introduce the flash temperature $T_\mathrm{f}$ and the critical temperature $T_\mathrm{c}$ as the temperature satisfying Eqs.~\eqref{eq:15} and \eqref{eq:16}, respectively.
Physically, $T_\mathrm{f}$ can be regarded as the highest temperature at which the system can be self-bound, since positive $P$ is always obtained above $T_\mathrm{f}$.
On the other hand, $T_\mathrm{c}$ is the second-order phase transition point where the local minimum of $P$ disappears.

We briefly note that our calculation is done under 
the condition
    $\mu/\omega_\perp\ll 1$ and $T/\omega_\perp\ll 1$
    to neglect the effects of the transverse excited states.

\section{Results and discussion}\label{sec:3}

\subsection{Zero-temperature limit}
Before moving to the finite-temperature calculation, we show how our formalism is reduced to the zero-temperature EOS reported in Ref.~\cite{PhysRevA.88.033611}, to be self-contained.
At $T=0$, we replace the Fermi-Dirac distribution with the step function
\begin{align}
    f(\xi_{k}+\Sigma_{k}^{\rm HF})=\theta(k_{\rm F}-k),
\end{align}
where $k_{\rm F}$ is the Fermi momentum associated with the particle number as $N = {k_\text{F}}/{\pi}$.
For the self-energy,
we obtain
\begin{align}
    \Sigma_{k}^{\rm HF}
    =\frac{{L}}{2\pi}\int_{-k_{\rm F}}^{k_{\rm F}}dq\,
    [V_{dd}(0)-V_{dd}(q-k)],
    \label{eq:zero_temp_self_energy}
\end{align}
where we replaced the momentum summation with the integration as{$\sum_{k}\rightarrow\frac{L}{2\pi}\int_{-\infty}^{\infty}dk$}.
Incidentally, at $k\rightarrow\infty$, the self-energy is reduced to the Hartree shift as $\Sigma_{k\rightarrow\infty}^{\mathrm{HF}}\to N V_{dd}(0)$ since $\lim_{k\rightarrow\infty}V_{dd}(q-k) =  0$.
Using {Eqs.~\eqref{eq:internal_energy}, \eqref{eq:10} and \eqref{eq:zero_temp_self_energy}}, we obtain the energy per particle at $T=0$ {
\begin{align}
    \frac{E}{N}
    =\frac{k_{\rm F}^2}{6m}
    +\frac{\pi}{2k_{\rm F}}
    \int_{-k_{\rm F}}^{k_{\rm F}}\frac{dk}{2\pi}
    \int_{-k_{\rm F}}^{k_{\rm F}}\frac{dq}{2\pi}\,
    [V_{dd}(0)-V_{dd}(q-k)],
\end{align}}
which is equivalent to the equation in Ref.~\cite{PhysRevA.88.033611}.
The numerical results at $\alpha\ell_{dd}/\ell_\perp=1.44$ and $1.90$ are shown in Fig.~\ref{fig:1}.
The equilibrium density (also called saturation density)
given by 
\begin{align}
    \left.\frac{\partial (E/N)}{\partial n}\right|_{n=n_0}=0
\end{align}
is found to be $n_0\ell_\perp\approx 0.31$ numerically in the present system at $T=0$ and $\alpha\ell_{dd}/\ell_\perp=1.9$, being consistent with the previous work~\cite{PhysRevA.88.033611}. In the following sections, we consider only the interaction strength for which the self-bound state is formed at $T=0$, namely $\alpha \ell_{dd}/\ell_\perp \gtrsim 1.8.$

One can obtain $P$ at $T=0$ from Eq.~\eqref{eq:14} as
\begin{align}
    P = 
    -{\frac{1}{L}}\int_{-\infty}^{\infty}\frac{dk}{2\pi}\left( \xi_k+\frac{1}{2}\Sigma_k^{\rm HF} \right)\theta\left( k_{\rm F}-k \right).
\end{align}
Meanwhile, $\mu$ at $T=0$ is obtained as
\begin{align}
    \mu = \frac{k_\mathrm{F}^2}{2m} +\Sigma_{k_\mathrm{F}}^{\mathrm{HF}}.
\end{align}
Using this, one can find
\begin{align}
\begin{split}
    P
    &= \frac{\pi^2n^3}{3m}+\frac{1}{\pi{L}}\int_0^{k_\mathrm{F}}dk\left(\Sigma_{k_\mathrm{F}}^{\mathrm{HF}}-\frac{1}{2}\Sigma_{k}^{\mathrm{HF}}\right) .
    \end{split}
    \label{eq:3.6}
\end{align}
While the first term in Eq.~\eqref{eq:3.6} is the pressure of an ideal Fermi gas at $T=0$, the second term is the interaction correction.

\subsection{Finite-temperature equation of state and phase diagram}
\begin{figure*}[t]
  \centering
  \includegraphics[width=0.95\linewidth]{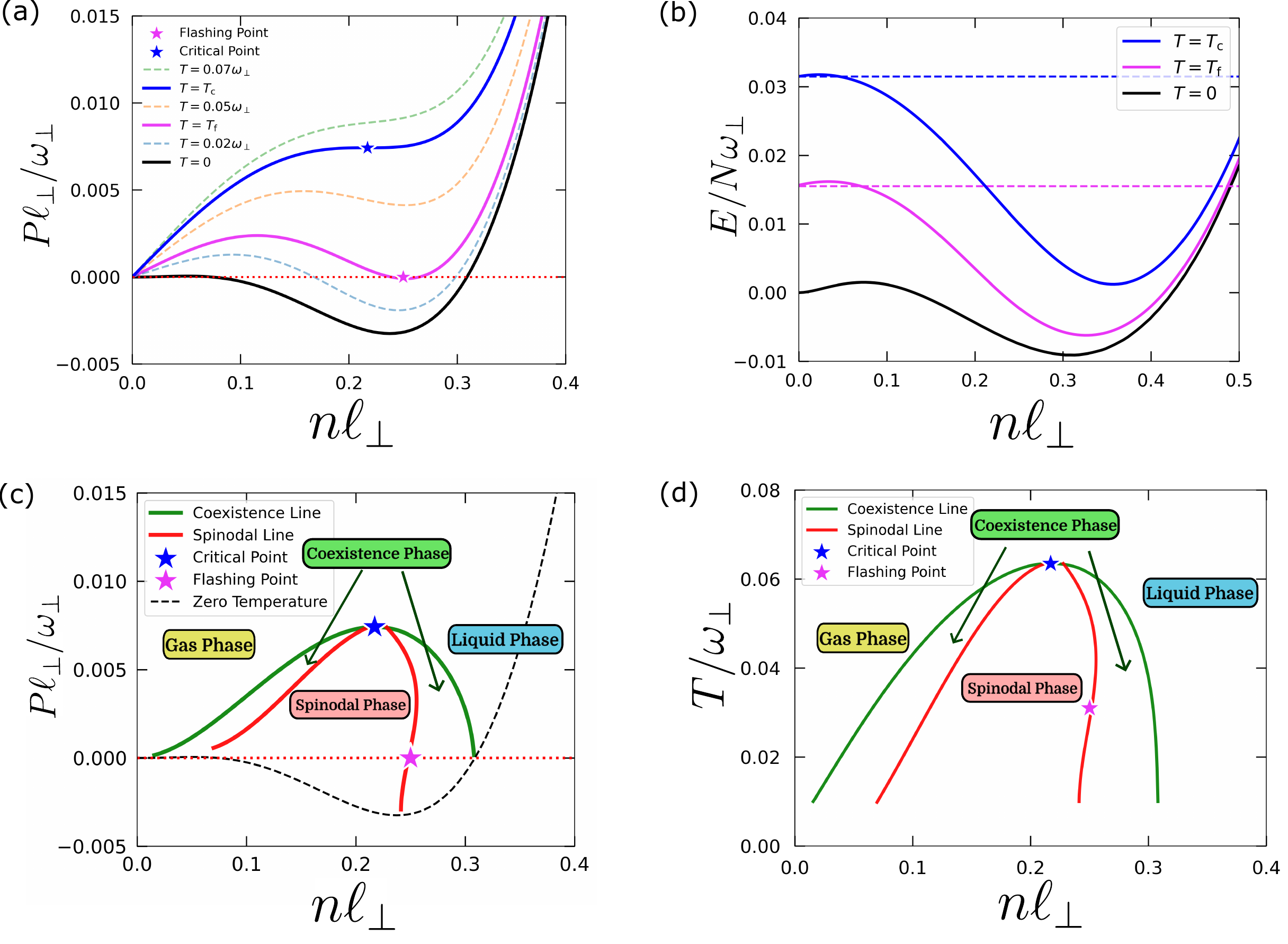}
  \caption{(a) Pressure $P$ and (b) energy per particle $E/N$ at several temperatures with $\alpha \ell_{dd}/\ell_\perp =1.9$. The black solid line represents the zero-temperature results. The blue and purple lines represent isotherms at critical temperature and flash temperature in the panels (a) and (b).
  The two dashed lines with light blue and yellow colors in panel (a) represent the isotherms at $T/\omega_\perp = 0.02 $ and $0.07$, respectively. Also, the red dotted horizontal line in panel (a) represents $P= 0$, which is tangent to the flash temperature isotherm.
  In panel (b), the two horizontal dotted lines show $E/N\omega_\perp = T_\mathrm{c}/2\omega_\perp=0.0315 $ and $E/N\omega_\perp = T_\mathrm{f}/2\omega_\perp =  0.0155$, respectively, showing that $E/N\omega_\perp = T/2\omega_\perp$ at the dilute limit.
  (c) Phase diagram with respect to pressure and particle density with $\alpha\ell_{dd}/\ell_\perp = 1.9$. The spinodal line satisfies $\partial P/\partial n=0$, and the coexistence curve corresponds to states with the same $T$, $P$, and $\mu$. The black dashed line represents EOS at zero temperature, which is the lower limit of pressure for finite temperature. (d) Temperature-density ($T-n$) phase diagram.
  The left and right regions with respect to the coexistence line represent the gas phase and the liquid phase, respectively.}
  \label{fig:phase_dia}
\end{figure*}
In this section, we present the phase diagram obtained from the self-consistent solution within the Hartree-Fock approximation.
In the numerical calculation, first we obtain the self-energy from the self-consistent equation~\eqref{eq:12} for a given number density $n$; thereafter, other thermodynamic quantities are calculated via thermodynamic identities. 

Figure ~\ref{fig:phase_dia}(a) shows the equations of state in terms of the $P$–$n$ relation at several temperatures: $T =0$, $0.02\omega_\perp$, $T_\mathrm{f}$, $0.05\omega_\perp$, $T_\mathrm{c}$, and $0.07\omega_\perp$, for $\alpha\ell_{dd}/\ell_\perp = 1.90$.
It can be seen that the pressure increases monotonically with temperature.  
At $T=0$, the system exhibits a self-bound state, and thus the pressure becomes negative below a certain density.  
As the temperature increases, the minimum pressure gradually increases, and the pressure first becomes non-negative at the flash temperature $T_{\mathrm{f}}$.  
Furthermore, with increasing temperature, the downward-convex region of the isotherm continuously shrinks and eventually disappears.  
The temperature at which this downward convex region vanishes corresponds to the critical temperature, $T_{\mathrm{c}}$.

Figure~\ref{fig:phase_dia}(b) shows the energy per particle $E/N$, as a function of $n$ at $T = 0$, $T_\mathrm{f}$, and $T_\mathrm{c}$. As the system exhibits a self-bound state with a local minimum of $E/N$ at $T = 0$, one can observe that the local minimum of $E/N$ occurs at nonzero low temperatures.
Note that the criterion for the formation of the self-bound liquid does not correspond to the presence of a local minimum in
$E/N$, in contrast to the zero-temperature case. 
In this regard, the thermodynamic potential $\Omega$, or $P$, should be considered for the determination of the critical and flash points.

Regarding the finite-temperature effect on $E/N$, one can find that $E/N$ monotonically increases with $T$ in Fig.~\ref{fig:phase_dia}(b).
In addition to the Fermi-degenerate pressure and the interaction correction, 
thermal contributions can be found at nonzero temperatures.
In particular, even at the dilute limit, $E/N$ approaches nonzero positive values in contrast to the zero-temperature case.
Indeed, at the dilute limit, the momentum distribution can be approximated by the Boltzmann distribution,
in which the particle number density and the energy density are given by
\begin{align}
    N&=\frac{z{L}}{\Xi}\int_{-\infty}^{\infty} \frac{dk}{2\pi}e^{-\frac{k^2}{2mT}}
    +O(z^2),\\
    E&=\frac{z{L}}{\Xi}\int_{-\infty}^{\infty} \frac{dk}{2\pi}\frac{k^2}{2m}e^{-\frac{k^2}{2mT}}
    +O(z^2),
\end{align}
where 
$z=e^{\beta\mu}$ is the fugacity.
The energy per particle in the classical limit is given by
\begin{align}
\label{eq:3.10}
    \frac{E}{N}=\frac{\int_{-\infty}^{\infty}\frac{dk}{2\pi}\frac{k^2}{2m}e^{-\frac{k^2}{2mT}}
    }{\int_{-\infty}^{\infty}\frac{dk}{2\pi}e^{-\frac{k^2}{2mT}}
    }
    =\frac{T}{2},
\end{align}
which is simply the thermal energy proportional to $T$.
This result enables us to confirm the convergence of the energy per particle found in Fig.~\ref{fig:phase_dia}(b).
It is also obvious from the equipartition theorem that one degree of freedom (i.e., spinless particles in one spatial dimension) corresponds to the kinetic energy of $T/2$.
We also note that the limit of $n\to0$ corresponds to the non-interacting dilute limit since the interaction effect is negligible there within the Hartree-Fock approximation.

\begin{figure}[t]
    \centering
    \includegraphics[width=1.0\linewidth]{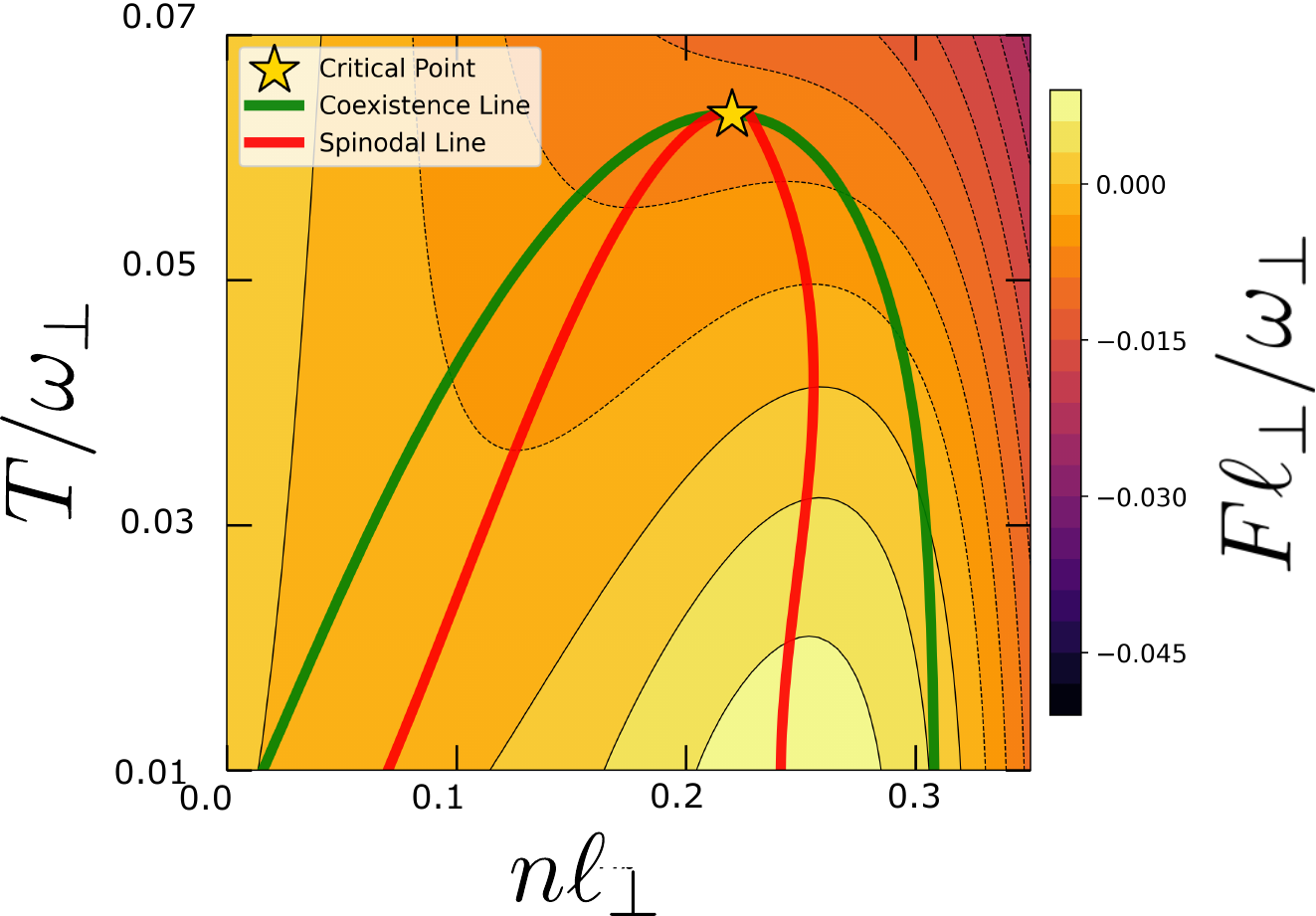}
    \caption{The free energy {density} $F$ in the plane of $T/\omega_\perp$ and $n\ell_\perp$.
    One can find a convex anomaly, in which the convex is upward, inside the coexistence phase (below the coexistence line). For comparison, we also show the spinodal line. }
    \label{fig:free_energy}
\end{figure}

Based on the thermal EOS, we could identify the critical point and the flash point given by Eqs~\eqref{eq:15} and \eqref{eq:16}; we can then draw the phase diagram. Figures~\ref{fig:phase_dia}(c) and (d) show the phase diagram of the system in the planes of $P-n$ and $T-n$, respectively.
The coexistence region lies inside the coexistence line, on which two thermodynamic states (i.e., the liquid and the gas) share the same values of the intensive variables:
\begin{align}
    P_\mathrm{gas} = P_\mathrm{liquid}, \,\,\,\,T_\mathrm{gas} = T_\mathrm{liquid}, \,\,\,\, \mu_\mathrm{gas} = \mu_\mathrm{liquid}.
\end{align}
We determine the coexistence line numerically by locating isothermal points at which both phases exhibit identical pressure and chemical potential.
The liquid and gas states on the coexistence line correspond to the high-density and low-density branches, respectively, which can thermodynamically coexist. This thermodynamic coexistence signals a first-order phase transition along the coexistence line.
  Figure~\ref{fig:free_energy} shows that the free energy {density} $F=-P+\mu n$ obtained by the Hartree-Fock approximation exhibits a convex anomaly in the coexistence phase, where $F$ is normally convex upward with respect to its natural variables $n$ and $T$.
  This is a clear feature of the liquid-gas phase transition.
  Note that a homogeneous miscible phase below the coexistence line obtained by the Hartree-Fock approximation also exists as a metastable thermodynamic state, which is different from the coexistence state. 

The spinodal line is defined by 
\begin{align}
    \left(\frac{\partial P}{\partial n}\right)_T = 0.
\end{align}
Below the spinodal line, the system undergoes the spinodal {instability}, where
$\partial P/\partial n<0$, or $\partial^2 F/\partial n^2<0$.
In this region, the homogeneous phase exhibits a thermodynamic instability 
with a spontaneous growth of density fluctuations
resulting in the phase separation.
This phenomenon is called spinodal decomposition,
which is a main origin 
of multi-fragmentation in nuclear matter~\cite{Chomaz2004Spinodal}.
This is also related to the convexity anomalies of the thermodynamic potential as shown in Fig.~\ref{fig:free_energy}.
In this way, the present cold-atomic system can mimic multi-fragmentation physics in nuclear matter.

{We note that the spinodal instability with vanishing wave vector $q=0$ may be preceded by an instability at a finite wave vector $q\neq 0$, leading to an inhomogeneous state.
Clarifying this possibility requires an analysis of the density-response function, for example, within the random phase approximation, and is left for future work.}

\subsection{Comparison with phenomenological models}
\begin{figure*}[t]
    \centering
    \includegraphics[width=0.95\linewidth]{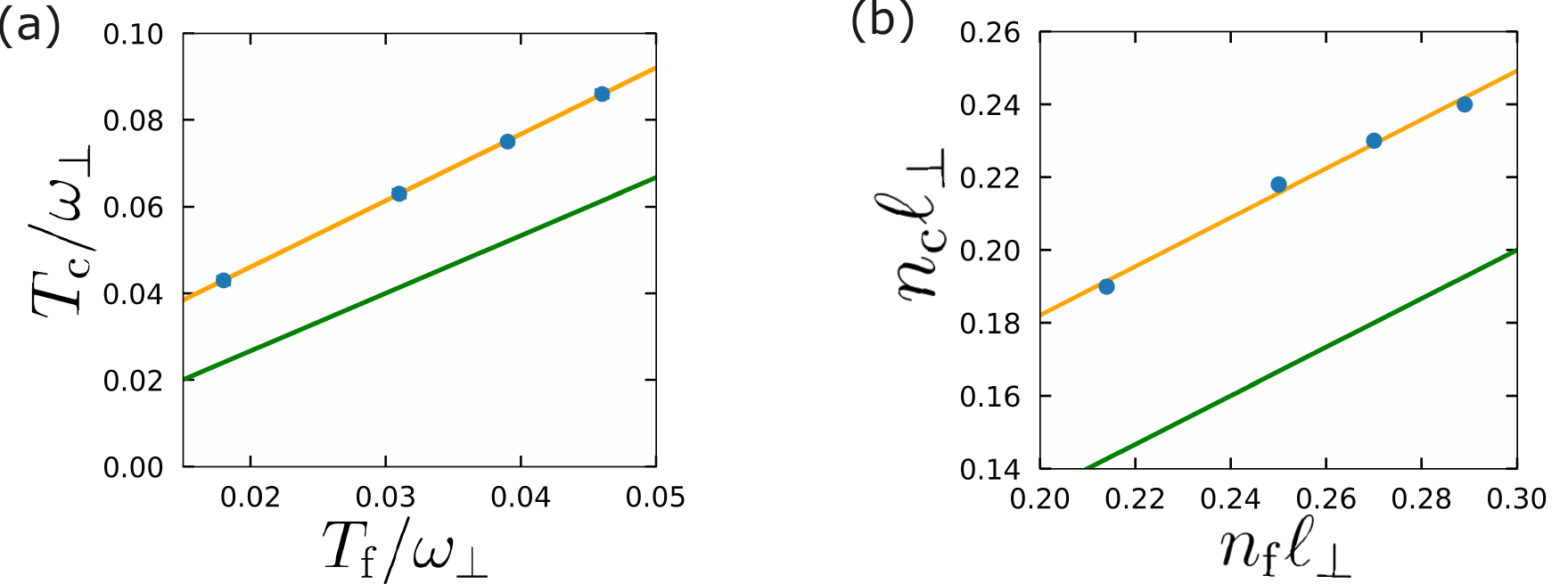}
    \caption{(a) The relationship between $T_\mathrm{c}$ and $T_\mathrm{f}$
    and (b) that between $n_\mathrm{c}$ and $n_\mathrm{f}$
    with different interaction strength $\alpha \ell_{dd}/\ell_\perp$ = 1.8, 1.9, 1.95, and 2.0 (from left to right). The blue dots show the numerical data of the Hartree-Fock approximation and the orange solid line shows the linear fit given by $T_\mathrm{c} = 1.53 T_\mathrm{f} +0.015\omega_\perp$ and
    $n_\mathrm{c}\ell_\perp = 0.67n_\mathrm{f}\ell_\perp + 0.048$.
    The green lines show the results of the Jaqaman model given by $T_\mathrm{c}/T_\mathrm{f} = 4/3.$ and $n_\mathrm{c}/n_\mathrm{f} = 2/3$, respectively.
    }
    \label{fig:TcTf}
\end{figure*}

Here, we discuss the flash and critical points in more detail.
The relation between critical points and flash points with different interaction strengths, $\alpha \ell_{dd}/\ell_\perp = 1.8, 1.9, 1.95$, and  $2.0$, is shown in Fig.~\ref{fig:TcTf}.
One can see that our results show a linear relation given by $T_\mathrm{c}=1.53T_\mathrm{f}+0.015\omega_\perp$ and $n_\mathrm{c}\ell_\perp = 0.67n_\mathrm{f}\ell_\perp + 0.048$.
Note that $\alpha \ell_{dd}/\ell_\perp > 2.0$ is not shown in Fig.~\ref{fig:TcTf} since the transverse excitation becomes non-negligible in this regime. 

To get the physical insight into this result, we compare our result with two effective models used in nuclear systems: the Jaqaman model and the Kapusta model \cite{Jaqaman1983,Rios2010phase,PhysRevC.29.1735}.

First, we adopt the Jaqaman model with the pressure
\begin{align}
    P_{\rm J} = nT + A n^2(n-n_0),\label{eq:29}
\end{align}
where $n_0$ and $A$ are parameters. 
In particular, $n_0$ corresponds to the saturation density.
At $n=n_0$ and $T=0$, $E/N$ exhibits a minimal value, and equivalently, $P$ vanishes at zero temperature. The second term in Eq.\eqref{eq:29} is taken to be a cubic function of $n$,
which is similar to the Skyrme force used in nuclear systems~\cite{Rios2010phase}. Thus, the term is essentially the expansion of energy around saturation density. 
On the other hand, the thermal contribution (i.e., temperature-dependent part) follows the equation of state in an ideal Boltzmann gas (see also Eq.~\eqref{eq:3.10}).
In this sense, this model considers only thermal fluctuations in a simple manner
and neglects quantum fluctuations.
Note that the temperature-dependent part is also applicable in a one-dimensional system, even though the original formula was developed for a three-dimensional system, because the equation of state of the ideal gas holds the same form across different dimensions.
Using the conditions of the flash point and the critical point given by Eqs.~\eqref{eq:15} and \eqref{eq:16},
we obtain 
\begin{align}
    T_\mathrm{f} = \frac{An_0^2}{4},\quad n_\mathrm{f} = \frac{n_0}{2}, \quad T_\mathrm{c} = \frac{An_0^2}{3}, \quad n_\mathrm{c} = \frac{n_0}{3}.
\end{align}
One can see that $T_\mathrm{c}/T_\mathrm{f} = 4/3$ and $n_\mathrm{c}/n_\mathrm{f} = 2/3$ do not depend on any parameters of the system, namely, $A$ and $n_0$.
These linear behaviors qualitatively agree with our Hartree-Fock results $T_\mathrm{c}=1.53T_\mathrm{f}+0.015\omega_\perp$ and $n_\mathrm{c}\ell_\perp = 0.67n_\mathrm{f}\ell_\perp + 0.048$, indicating that our results can partially be understood as classical thermal fluctuations on top of the ground-state equation of state.

Second, we consider the Kapusta model, which is also often used in understanding $P-n$ relation in nuclear matter \cite{PhysRevC.29.1735,Rios2010phase}.
In contrast to the Jaqaman model with classical thermal fluctuations,
thermal excitation on top of the non-interacting Fermi-degenerate state is taken into account in the Kapusta model. 
In three-dimensional systems, its equation of state is expressed as
\begin{align}
    P_{\rm K}^{\rm 3D} = An^2(n-n_0) + B n^{1/3}T^2,
\end{align}
where $A$ and $B$ are parameters.
Here, $B$ is proportional to the effective mass $m^*$ at the saturation density. The temperature-dependent part is proportional to $T^2$. 
This can be derived by using the Sommerfeld expansion,
which, however, depends on spatial dimensions.
In this regard, we have to generalize it to the one-dimensional case to compare it with our results.

To this end, we start from the finite-temperature expression of the internal energy {per unit length}
\begin{align}
\label{eq:3.16}
    E_\mathrm{K} = \int_0^\infty d\varepsilon\,\, \varepsilon D(\varepsilon) f(\varepsilon),
\end{align}
where $D(\epsilon) = \frac{\sqrt{m^*}}{\pi\sqrt{2}}\frac{1}{\sqrt{\varepsilon}}=\mathscr{D}/\sqrt{\varepsilon}$ is the density of states (DOS) per unit length in a one-dimensional system.
Since we are interested in the region near $n=n_0$, 
we may consider the effective mass $m^*$ at $n=n_0$ and $T=0$.
For convenience, we have introduced a constant $\mathscr{D}=\frac{\sqrt{m^*}}{\pi\sqrt{2}}$.
Applying the Sommerfeld expansion to the energy integral in Eq.~\eqref{eq:3.16},
we get
\begin{align}
    E_\mathrm{K} = \int_{0}^\mu \mathscr{D}\sqrt{\varepsilon} d \varepsilon + 
    \frac{\pi^2\mathscr{D}}{12}\frac{T^2}{\sqrt{\mu}}
    +O(T^3).
\end{align}
The chemical potential can also be expanded by the Sommerfeld expansion as $\mu = \varepsilon_\mathrm{\mathrm{F}} + \frac{\pi^2}{12} \frac{T^2}{\varepsilon_{\rm F}}+O(T^3)$, where $\varepsilon_{\rm F}$ is the Fermi energy. 
Using the thermodynamic relation $(\partial S/\partial T)_V = (\partial E/\partial T)_V /T$, 
we obtain the entropy {per unit length} in the {1D} Kapusta model as
\begin{align}
    S_{\rm K}^{\rm 1D} = \frac{m^*Tn^{-1}}{3} +O(T^2)
\end{align}
in the Fermi-degenerate limit of a one-dimensional system. 
This suggests that, in the Fermi-degenerate limit (in other words, in the high-density limit), the entropy is proportional to $T/T_{\mathrm{F}}\propto m^*Tn^{-2}$, where $T_\mathrm{F}$ is the Fermi temperature.
From the thermodynamic relation, one can obtain $P = n^2(\partial(E/N)/\partial n)_{s}$ with fixed entropy per particle $s\equiv {L}S_{\rm K}^{\rm 1D}/N = m^*Tn^{-2}/3$. 
In this way, the temperature-dependent part of the pressure is proportional to $m^*T^2n^{-1}$, and we have the equation of state in the one-dimensional Kapusta model as
\begin{align}
    P_{\rm K}^{\rm 1D}=An^2(n-n_0) + \frac{m^*}{3n}T^2. \label{eq:Kapusta}
\end{align}

This model accurately describes the behavior in the Fermi-degenerate region. For example, we obtain $A/\omega_\perp\ell_\perp^2 \approx 1.5$ and $n_0\ell_\perp \approx 0.31$ by the fitting procedure. Then, we obtain $m^*/m \approx 1.8$ at $n = n_0$ from Fig.~\ref{fig:eff_mass}.
The calculation of effective mass can be found in Appendix A. 
From Eq.~\eqref{eq:Kapusta}, we obtain $T_\mathrm{f} = \sqrt{\frac{A n_\mathrm{f}n_0}{3m^*}}$ using $P =0$ and $dP/dn = 0$. We can find the flash point in the Kapusta model as 
\begin{align}
T_\mathrm{f}/\omega_\perp \approx 0.028,\quad n_\mathrm{f}\ell_\perp \approx 0.23    
\end{align}
at $\alpha\ell_{dd}/\ell_\perp = 1.9$. This result agrees with the self-consistent calculation of the Hartree-Fock theory where $T_\mathrm{f}/\omega_\perp \approx 0.031$, $n_\mathrm{f} \ell_\perp \approx 0.25$, indicating that our main result around the flash point can be well understood in terms of the Kapusta model with the effective mass at saturation density.

\begin{figure}
    \centering
    \includegraphics[width=0.95\linewidth]{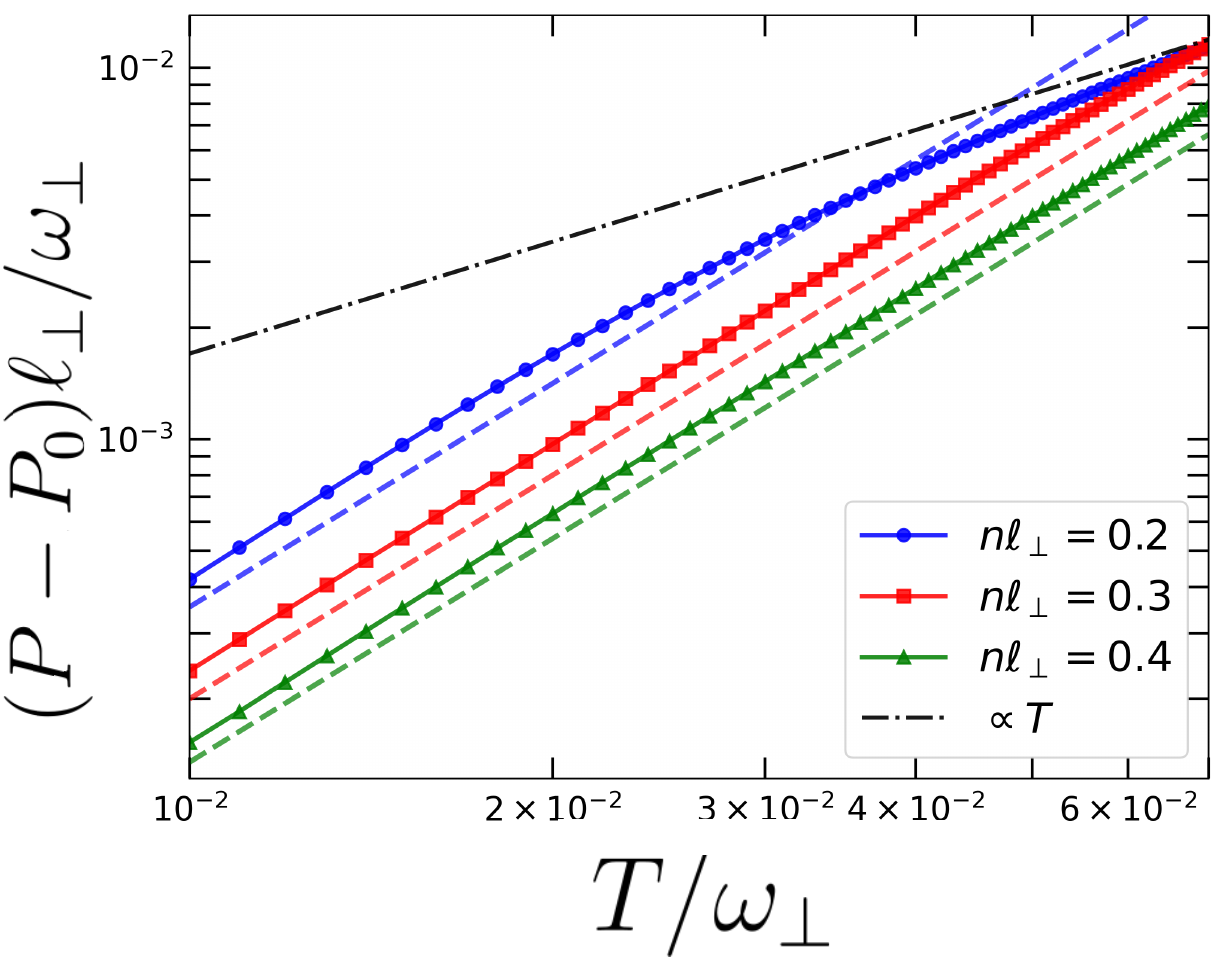}
    \caption{Scaling of thermal effect of pressure $P$. Here, $P_0$ represents the pressure at $T=0$. It can be seen that higher density ($n\ell_\perp = 0.3$ and $0.4$ in this case) exhibits a quadratic dependence on $T$ while the behavior at $n\ell_\perp = 0.2$ becomes linear as $T$ increases at higher temperature (where the chain-dotted line represents $\propto T$). The dash line represents the extended Kapusta model with density-dependent effective mass, where the $T$ dependence of the thermal pressure is quadratic.
    }
    \label{fig:TvsT2}
\end{figure}

Figure~\ref{fig:TvsT2} represents how $P$ deviates from the zero-temperature case denoted by $P_0$ at finite temperature. 
One can see in Fig.~\ref{fig:TvsT2} that the Jaqaman model captures the temperature dependence of the pressure at densities below saturation density and at high temperature, while the Kapusta model gives a better description in the vicinity of and beyond the saturation density at low temperatures.
These correspond to the regions in which the classical-ideal-gas approximation and the Fermi-degenerate limit make sense, respectively.
The classical-ideal-gas approximation is valid when the system is dilute so that 
the Fermi distribution is reduced to the Boltzmann distribution due to the high temperature compared to the Fermi temperature $T_\mathrm{F} = n^2\pi^2/2m^* \ll T$ and the negligible interaction (i.e., $m^*\simeq m$).
In contrast, the Fermi-degenerate limit is realized when $T_\mathrm{F} \equiv n^2\pi^2/2m^* \gg T$. 
In this sense, the Kapusta model can be regarded as the opposite limit in terms of the number density compared to the Jaqaman model.

Meanwhile, both the Jaqaman and Kapusta models cannot be used at the non-classical dilute limit (i.e., $n\rightarrow 0$ and $T\rightarrow 0$ with $n\lambda_T>1$ where $\lambda_T=\sqrt{2\pi/mT}$ is the thermal de Broglie length) because the one-dimensional DOS at the Fermi level diverges near the quantum phase transition from finite density to vacuum.
Besides, in contrast to the Jaqaman model and three-dimensional Kapusta model, the one-dimensional Kapusta model does not yield a critical point due to the divergent behavior of pressure in $n\to 0$. This divergence is associated with the divergent DOS at low energies, which induces a large amount of thermal excitations at low densities.

\begin{figure}
    \centering
    \includegraphics[width=1.0\linewidth]{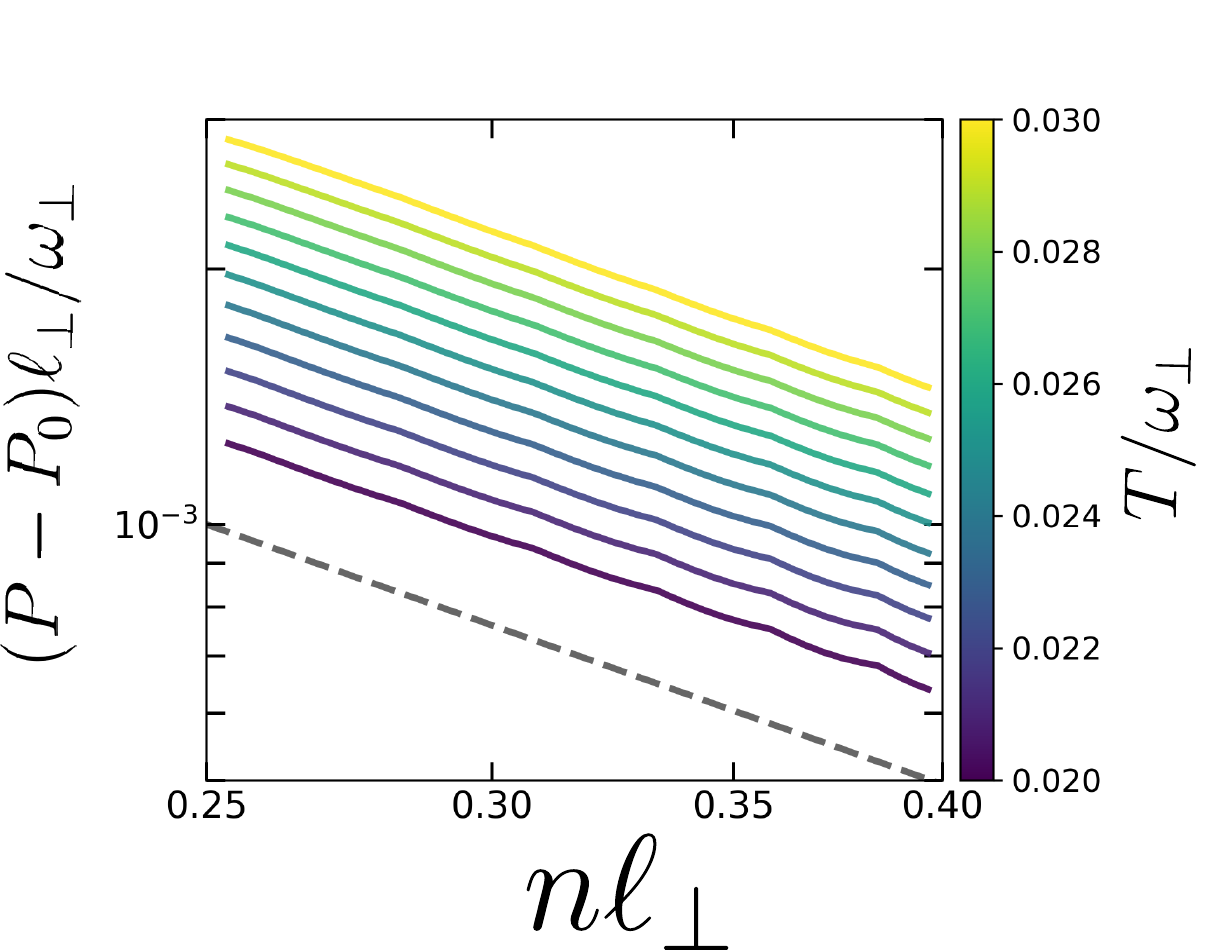}
    \caption{Log-log plot of
    the thermal contribution on the pressure $P-P_0$ as a function of $n$,  with different temperatures from $T/\omega_\perp = 0.02$ to 0.03. The dashed line shows the case of power $-1.5$ (i.e., $\propto n^{-1.5}$) for reference.}
    \label{fig:scalinngvsn_of_thermal_pressure}
\end{figure}

A more accurate effective model is to incorporate the density dependence of effective mass in the Kapusta model \cite{Rios2010phase}.
This is equivalent to replacing $m^*$ at $n=n_0$ in Eq.~\eqref{eq:Kapusta} with the density-dependent one $m^*(n)$ at $T=0$ shown in  Appendix~\ref{app:a}.
Since we focus on the EOS around saturation density $n_0$,
the Kapusta model is based on the expansion of EOS around $n=n_0$ with fixed $m^*(n=n_0)$.
In this sense, its valid region is limited to the region near $n=n_0$, that is, from $n\ell_\perp = 0.25 $ to 0.4, in which the temperature contribution to the pressure is well explained by $T^2$.
In this region, we see $P -P_0 \propto T^2/n^j$ where $j \approx 1.5$ as shown in Fig.~\ref{fig:scalinngvsn_of_thermal_pressure}.
Indeed, the power $j\approx 1.5$ in the thermal contribution differs by $\sim 0.5$ from Eq.~\eqref{eq:Kapusta} with $j=1$.
This discrepancy can be partially attributed to the density dependence of effective mass, which is approximately $m^*(n)\propto n^{-0.4}$ as shown in Fig.~\ref{fig:eff_mass}. 
Although the resulting power is different by a factor $0.1$, this small difference would come from the higher-order terms dropped in the expansion of $\Sigma_{k}^{\rm HF}$ around $k=k_{\rm F}$.
In this way, the extended Kapusta model gives us a more accurate description of the present system at finite temperature, in terms of the density-dependent effective mass.

\begin{figure*}[t]
    \centering
    \includegraphics[width=0.95\linewidth]{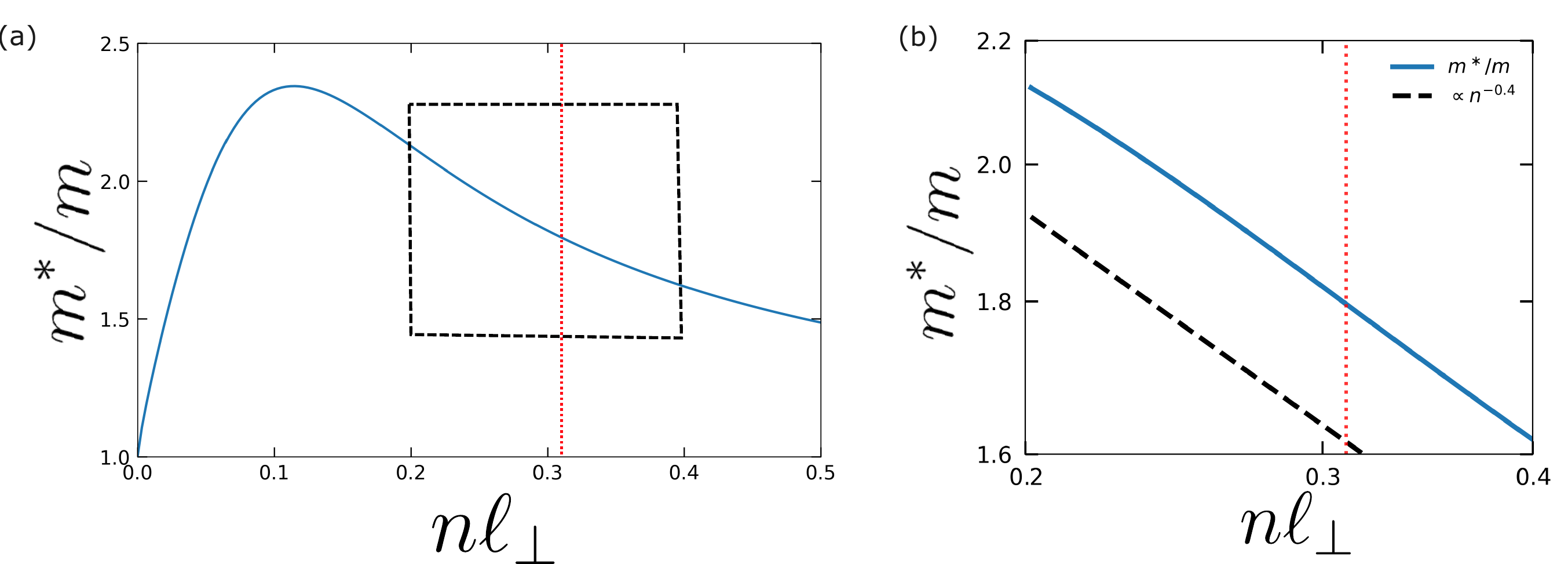}
    \caption{(a) Effective mass $m^*$ as a function of the number density $n$ at $\alpha\ell_{dd}/\ell_\perp = 1.9$ and $T=0$ within the Hartree-Fock approximation.
    The panel (b) is the log-log plot of $m^*/m$ around $n\ell_\perp = n_0\ell_\perp\approx 0.31$ (vertical dotted red line), indicating that $m^*/m$ is approximately proportional to $n^{-0.4}$ (dashed line) in this regime.}
    \label{fig:eff_mass}
\end{figure*}

\subsection{Discussion}
Although the Hartree-Fock theory qualitatively agrees with the result of generalized Bose-Fermi duality at $T=0$~\cite{PhysRevA.88.033611}, we briefly remark on properties of phase transitions and limitations in our approach. 
The present interaction potential in real space is given by~\cite{PhysRevA.88.033611,PhysRevLett.99.140406}
\begin{align}
\begin{split}
    V(z) &= -\frac{\alpha d^2}{\ell_\perp^3}\int_0^\infty dx x^2e^{-x^2/2-x|z|/\ell_\perp} \\&= \frac{\alpha d^2}{\ell_\perp^3}\left[\frac{|z|}{\ell_\perp}-\sqrt{\frac{\pi}{2}}e^{\frac{|z|^2}{2\ell_\perp^2}}\left(1+\frac{|z|^2}{\ell_\perp^2}\right)\mathrm{erfc\left(\frac{|z|}{\sqrt{2}\ell_\perp}\right)}\right],
    \end{split}
\end{align}
where ${\rm erfc}(x)$ is the complementary error function.
The potential in 1D behaves asymptotically as $V(z) \sim -1/|z|^3$ when $|z| \gg \ell_\perp$ and $V(z) \sim (-\sqrt{\pi /2} +|z|/\ell_\perp)$ when $|z| \ll \ell_\perp$. Therefore, it can be said that $V(z)$ is a long-range interaction on a short-distance scale while it acts as a short-range interaction on a long-distance scale. Note that  $1/|z|^3$  in 1D systems is treated as a short-range interaction in statistical mechanics because it conserves the additivity of energy in the thermodynamic limit~\cite{Mukamel2009_LongRangeNotes}.

Although it is known that 1D systems with short-range interactions are prohibited from exhibiting continuous phase transitions or spontaneous continuous symmetry breakings~\cite{MerminWagner1966,PhysRev.158.383}, it is possible to observe a first-order phase transition with long-range interactions. 
In this system, this is the case due to long-range-like behavior of interaction in the typical particle separation $\ell_\perp$. 

However, we should note that our result is totally limited by the Hartree-Fock approximation. It is known that mean-field calculations may lead to a phase transition that does not actually happen, e.g., the 1D Ising model. {In particular, in 1D systems, fluctuations can be significant and we have to admit that the phase transition might not be observed if we take higher orders into account.}
To consider the legitimacy of the result, we should note that the Hartree-Fock approximation is a good approximation in high-density regions {i.e., at the densities much larger than the saturation density}, where kinetic energy dominates, and that the energy obtained in the Hartree-Fock approximation is an upper bound of the exact energy. 

{As we mentioned above, interacting fermions are generally treated as the TLL in one-dimensional systems as a low-energy effective theory near the Fermi surface since it is a strongly correlated system. Thus, the Hartree-Fock approximation is not quantitatively accurate in the low-density region. However, as verified in \cite{PhysRevA.88.033611}, the Hartree-Fock method can nonetheless reproduce the self-bound state at $T=0$ in this density region, indicating that the Hartree-Fock method is at least qualitatively valid in the low-density region to discuss the liquid-gas phase transition. We also note that the TLL picture is not valid when the Fermi energy is comparable to the temperature. Indeed, near the critical point, the Fermi temperature $T_F/\omega_\perp = \pi^2n^2/2m^* \approx 0.3$ while $T_c/\omega_\perp \approx 0.06$, indicating that the TLL picture is not valid near the critical point. In this sense, the Hartree-Fock method is a reasonable starting point to discuss the liquid-gas phase transition in the present system.}

{
\begin{table*}[t]
\centering
\caption{Order-of-magnitude experimental parameters for realizing the strong-coupling regime $\ell_{dd}/\ell_\perp \sim 2$. Here, $\ell_{dd}$ is defined in our convention. Here, $\mu_B$ is the Bohr magneton.}
\begin{tabular}{lcccccc}
\hline\hline
Species & Dipole moment  & $\ell_{dd}$ & $\ell_\perp$ required & $n_c$ & $\omega_\perp/2\pi$ & $T_c$ \\
 &  &  & for $\ell_{dd}/\ell_\perp \sim 2$ &  &  & $(T_c/\omega_\perp\sim 0.06)$ \\
\hline
$^{23}$Na$^{40}$K & $d \simeq 0.8~\mathrm{D}$ & $\sim 0.6~\mu\mathrm{m}$ & $\sim 0.3~\mu\mathrm{m}$ & $\sim 10^4~\mathrm{cm}^{-1}$ & $\sim 2~\mathrm{kHz}$ & $\sim 6~\mathrm{nK}$ \\
$^{40}$K$^{87}$Rb & $d \simeq 0.2~\mathrm{D}$ & $\sim 0.08~\mu\mathrm{m}$ & $\sim 0.04~\mu\mathrm{m}$ & $\sim 10^5~\mathrm{cm}^{-1}$ & $\sim 50~\mathrm{kHz}$ & $\sim 140~\mathrm{nK}$ \\
$^{161}$Dy & $\mu \simeq 10\,\mu_B$ & $\sim 20~\mathrm{nm}$ & $\sim 10~\mathrm{nm}$ & --- & sub-MHz to MHz & --- \\
$^{167}$Er & $\mu \simeq 7\,\mu_B$ & $\sim 5~\mathrm{nm}$ & $\sim 10~\mathrm{nm}$ & --- & sub-MHz to MHz & --- \\
\hline\hline
\end{tabular}
\label{tab:exp_parameters}
\end{table*}}

{To realize this system, we briefly discuss the experimental parameters. Dipolar fermions are realized by cold polar molecules such as $^{23}$Na$^{40}$K and $^{40}$K$^{87}$Rb, and strongly magnetic atoms such as $^{161}$Dy and $^{167}$Er~\cite{biswas_controlled_2026,PhysRevLett.108.215301,Park2015NaKGroundState,deMiranda2011ultracold,Aikawa2014fermiDegeneracy,PhysRevLett.121.093602}. An important parameter for realization of this experiment is the condition that the transverse confinement length and the dipole length are comparable, i.e., $\ell_{dd}/\ell_\perp \sim 2.0$, to study this strong-coupling regime.

Ground-state $^{23}$Na$^{40}$K is chemically stable and induced dipole moments up to $d \simeq 0.8~\mathrm{D}$ have been demonstrated~\cite{Park2015NaKGroundState}. In our convention, this corresponds to a dipole length $\ell_{dd} \sim 0.6~\mu\mathrm{m}$. The regime studied in this work, $\ell_{dd}/\ell_\perp \sim 2$, therefore requires a transverse confinement length $\ell_\perp \sim 0.3~\mu\mathrm{m}$, which corresponds to a confinement frequency $\omega_\perp/2\pi \sim 2~\mathrm{kHz}$ for $^{23}$Na$^{40}$K. Then the critical density is estimated to be $n \sim 0.2/\ell_\perp \sim 10^4~\mathrm{cm}^{-1}$.

On the other hand, $^{40}$K$^{87}$Rb has a slightly smaller dipole moment, and $\ell_{dd}$ inducible in experiments is estimated to be $\ell_{dd}\sim 0.08~\mu\mathrm{m}$~\cite{Ni2010dipolar}, which requires a tighter confinement $\ell_\perp \sim 0.04~\mu\mathrm{m}$, corresponding to $\omega_\perp/2\pi \sim 50~\mathrm{kHz}$ for $^{40}$K$^{87}$Rb. The estimated density for the observation of the liquid-gas phase transition is $n \sim 0.2/\ell_\perp \sim 10^5~\mathrm{cm}^{-1}$.

For magnetic atoms such as $^{161}$Dy and $^{167}$Er, the dipole moment is magnetic and their $\ell_{dd}$ is much smaller than that of polar molecules (e.g., $\ell_{dd}\sim 20~\mathrm{nm}$ for $^{161}$Dy and $10~\mathrm{nm}$ for $^{167}$Er~\cite{PhysRevLett.108.215301}), which requires an even tighter confinement $\ell_\perp \sim 10~\mathrm{nm}$, corresponding to sub-MHz to MHz confinement frequencies.

It is worth noting that the observation of a fermionic self-bound droplet has been limited compared to its bosonic counterpart. In a single-component Fermi atom gas, efficient evaporative cooling and rethermalization for fermions are intrinsically more difficult than in bosonic gases, although dipolar scattering can partly restore elastic collisions for magnetic fermions~\cite{PhysRevA.88.033611,Aikawa2014fermiDegeneracy}. Besides, magnetic dipole of atoms are not strong enough to form a self-bound droplet, as we have seen in the discussion above. Moreover, the fermionic polar-molecule route relevant for the present proposal was limited by short-range reactive or inelastic losses~\cite{Ospelkaus2010quantumStateControlled,Schindewolf2022evaporation}. Indeed, the low-dimensional geometries that successfully suppressed KRb losses relied on repulsive side-by-side collisions~\cite{deMiranda2011ultracold}, whereas the present self-bound regime requires attractive head-to-tail intratube interactions. In this sense, only recently have the ingredients of deep degeneracy, strong dipolar interactions, and sufficient collisional stability begun to coexist in a single fermionic platform by leveraging microwave shielding~\cite{Schindewolf2022evaporation,biswas_controlled_2026}.

The remaining challenge is to combine such confinement with sufficient collisional stability and sufficient low temperature in a quasi-one-dimensional geometry with attractive intratube dipolar interactions. For the representative value $T_c/\omega_\perp \sim 0.06$ obtained in this work, the corresponding critical temperature is estimated to be in the $6~\mathrm{nK}$ range for NaK and $140~\mathrm{nK}$ for KRb; the latter is achievable for KRb in current experiments~\cite{biswas_controlled_2026,de2019degenerate}. The estimated parameters for realizing the strong-coupling regime $\ell_{dd}/\ell_\perp \sim 2$ are summarized in Table~\ref{tab:exp_parameters}.}

\section{Summary}
\label{sec:4}
In this paper, we have discussed the thermal liquid-gas phase transition in a quasi-one-dimensional single-component Fermi gas with the attractive dipole-dipole interaction.
Within the Hartree-Fock approximation, we have elucidated the occurrence of the first-order phase transition involving the coexistent liquid and gas phases and the spinodal instability at finite temperatures.
The calculated flash and critical points in the present system show qualitatively good agreement with those of the phenomenological models (Jaqaman model and Kapusta model) used in the study of liquid-gas phase transition in nuclear matter. 
We have also elucidated the importance of the effective mass in the thermal equation of state.
This result can be useful for future cold-atom experiments of dipolar fermions under low-dimensional confinement.

For future perspectives, it is worth investigating the beyond-mean-field effect on the equation of state and the phase transitions.
The equation of state in the present low-dimensional system can also be used as a testing ground for many-body theories such as the functional-renormalization-group-aided density functional theory~\cite{kemler2016formation,liang2018functional,PhysRevC.99.024302} and the complex Langevin method~\cite{PhysRevD.98.054507,rammelmuller2020pairing,PhysRevResearch.3.033180}.
A study on $p$-wave pairing effect is also an interesting future direction, which can be investigated by using the Hartree-Fock-Bogoliubov theory~\cite{ring2004nuclear}.
The three-body force is also important for simulating the nuclear equation of state, and can be studied by preparing a mixture with a medium bath which induces the exchange of medium excitations between fermions~\cite{tajima2025tunable}.

\acknowledgments
The authors thank T. Fukui and M. Horikoshi for useful discussions.
This work is partially supported by the JSPS KAKENHI under Grants No.~JP22K13981, No.~JP23K22429, and No.~JP26K07063.

\section*{Data availability}
The data that support the findings of this article are not publicly available. The data are available from the authors upon reasonable request.

\appendix
\section{Effective mass}
\label{app:a}

Expanding the HF self-energy with respect to the momentum $k$ around the Fermi momentum $k_{\rm F}$, one can obtain the effective mass $m^*$ at $T=0$ as
\begin{align}
    \xi_k+\Sigma_k^{\mathrm{HF}} \approx \frac{k_{\rm F}}{m^*}(k-k_{\rm F}),
\end{align}
where we used
\begin{align}
    \frac{k_{\rm F}^2}{2m} =\mu-\Sigma_{k_{\textrm{F}}}^{\rm HF},
\end{align}
resulting in
\begin{align}
    \frac{m^*}{m}=\frac{1}{1+\frac{m}{k_{\rm F}}\left.\frac{\partial\Sigma}{\partial k}\right|_{k= k_{\rm F}}}\label{eq:effective_mass}.
\end{align}
Note that, in the Hartree-Fock approximation, the renormalization factor reads $Z=[1 - \partial\Re\Sigma(\varepsilon,p)/\partial  \varepsilon]^{-1} = 1$ (where $\varepsilon$ is the single-particle energy) since there is no frequency dependence in the self-energy.
One can directly obtain the effective mass $m^*$ from Eq.~\eqref{eq:effective_mass} by using
\begin{align}
    \left.\frac{\partial \Sigma_{k}^{\rm HF}}{\partial k}\right|_{k = k_{\rm F}} &= \frac{1}{2\pi}\int_{-k_{\rm F}}^{k_{\rm F}}dq\,\, V'(q-k_{\rm F})\cr
    &= \frac{V_{dd}(0) -V_{dd}(2k_{\rm F})}{2\pi}.
\end{align}
Figure~\ref{fig:eff_mass} shows the effective mass calculated
in $\alpha\ell_{dd}/\ell_\perp = 1.9$ and
$T=0$.
The fact that $m^*$ is associated with $V(2k_{\rm F})$ is intuitively understood as the effect of an exchange interaction between particles with momenta of $k_{\rm F}$ and $-k_{\rm F}$ on the Fermi surface.
In this regard, one can see the peak structure of $m^*/m$ in the density dependence.

\bibliographystyle{apsrev4-1}
\bibliography{reference.bib}

\end{document}